\documentclass[12pt,a4paper]{article}
\usepackage{graphicx}

\begin{document}
\textwidth=135mm
 \textheight=200mm
\begin{center}
{\bfseries Investigation of avalanche photodiodes radiation hardness for baryonic matter studies}
\vskip 5mm
V.~Kushpil$^{a,\dag}$, V.~Mikhaylov$^{a,b,c,\ddag}$, V.P.~Ladygin$^{d}$, A.~Kugler$^a$, S.~Kushpil$^a$, 
O.~Svoboda$^a$, P.~Tlust\'y$^a$     
\vskip 5mm
{\small {\it $^a$ Nuclear Physics Institute, Academy of Sciences of the Czech Republic, 25068 \v Re\v z, 
Czech Republic}} \\
{\small {\it $^b$ Czech Technical University in Prague, Faculty of Nuclear Sciences and Physical Engineering, 16636 Prague, Czech Republic }} \\
{\small {\it $^c$ National Research Tomsk Polytechnic University, Department of
Electronics and Automation of Nuclear Plants, 634050 Tomsk, Russian Federation}} \\
{\small {\it $^d$ Joint Institute for Nuclear Research, 141980 Dubna, Russian Federation}} \\
{\small {\it $^\dag$ E-mail: kushpil@ujf.cas.cz }}\\
{\small {\it $^\ddag$ E-mail: mikhaylov@ujf.cas.cz}}\\
\end{center}

\vskip 5mm
\centerline{\bf Abstract}
Modern avalanche photodiodes (APDs) with high gain are good device candidates for light readout from detectors applied in relativistic heavy ion collisions experiments.  
The results of the investigations of the APDs properties from Zecotek, Ketek and Hamamatsu  manufacturers
after irradiation using secondary neutrons from cyclotron facility U120M at NPI of ASCR in 
\v Re\v z are presented. The results of the investigations can be used for the
design of the detectors for the experiments at NICA and FAIR.   
\vskip 10mm

\section{\label{sec:intro}Introduction}

The studies of the properties of nuclear matter under extreme density and temperature conditions are
the main subject of the relativistic heavy-ion collisions experiments  at Nuclotron-based Ion Collider fAcility (NICA) and at Facility for Antiproton and Ion Research (FAIR). The study of the dense baryonic matter at Nuclotron (BM@N project) \cite{bmn_CDR} is proposed as a first stage in the heavy-ion program at NICA \cite{nica}. 
The research program of BM@N project includes the studies of  the production of strange matter in heavy ion collisions at beam energies between 2 and 6 A$\cdot$GeV \cite{bmn_PoS}, in-medium effects for 
strange particles decaying in hadronic modes \cite{brat1},
hard probes and correlations \cite{vasiliev_npps2011}, spin and polarization effects
\cite{lambda, bmn_dspin2013, bmn_baldin2014}.
These studies will be complementary to the Compressed Baryonic Matter (CBM) project research program \cite{CBM_pb, cbm} for fixed target heavy ion collisions at FAIR in future. 

The main advantages of APDs are very compact sizes, low bias voltage, gain comparable to that for standard
photo-multiplier  tubes (PMTs), relative low price, insensitivity to magnetic field and absence of nuclear counter effect (due to the pixel structure). APDs have the following typical properties: pixel density about 
10$^4$ -- 2$\cdot$10$^4$/mm$^2$, size of 3$\times$3~mm$^2$, high dynamical range of 5--15000 ph.e., photon detection efficiency of $\sim$15\%, high counting rate of $\sim$10$^5$~Hz. 

The APDs are proposed as a main option for the light readout 
for Forward Wall Detector (FWD) at the BM@N \cite{bmn_CDR} based on the high granularity scintillation hodoscope.
FWD  will allow to measure the spectators and fragments 
for the precise reconstruction of the reaction plane in the semi-central events required for the measurements of flows, global polarizations of the hyperons etc. 
The application of such detector for heavy ion collisions at HADES is described elsewhere \cite{HADES}. 
Another application of APDs is the  Projectile Spectator Detector (PSD) for the CBM setup. 
PSD is a detector of non-interacting nucleons and fragments emitted at very low polar angles in forward direction in nucleus-nucleus collisions \cite{psd}.  It will be used to determine the collision centrality 
and the orientation of an event plane.  The PSD is a fully compensating modular lead-scintillator calorimeter, which provides very good and uniform energy resolution. The calorimeter comprises 
44 individual modules, each consisting of 60 lead/scintillator layers with a cross section of 20$\times$20 cm$^2$.

The radiation hardness of APDs, especially to neutrons, plays a significant role for the 
the detectors placed in the forward direction. For instance,  according to FLUKA simulation \cite{fluka} neutron 
flux near the beam hole might achieve 10$^{12}$ neutrons/cm$^2$ for beam energy 4 A$\cdot$GeV  and 2 months of CBM run at the beam rate 10$^8$ ions per second \cite{psd, resp}.  
In this paper the results of radiation hardness of the APDs  from Zecotek, Ketek and Hamamatsu  manufacturers 
to neutrons are presented.
For the proposed experiments it is necessary to separate signal from noise for cosmic muons, while resolution of individual photons is not so important.

\section{\label{sec:setup} Setup for neutron irradiation studies}

\begin{figure}[hbtp]
 \centering
  \resizebox{10cm}{!}{\includegraphics{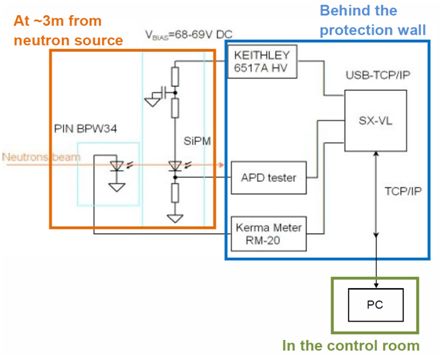}}
\caption{Setup for the beam fluence control and measurements of the APDs properties.}
\label{fig:fig1}
\end{figure}

The APDs  were irradiated  using quasi-monoenergetic 35 MeV secondary neutron beam from cyclotron facility U120M
at NPI of ASCR in  \v Re\v z \cite{cyclo}.  The schematic view of the setup for the beam fluence control and measurements of the irradiated  APDs properties is presented in Fig.~\ref{fig:fig1}. It consists of PIN diode BPW34 connected to Kerma Meter RM20 used for neutron fluence measurement \cite{kerma}, APD sample biased by voltage power supply from Keitheley 6517A,  APD tester \cite{apd_tester} and Tektronix oscilloscope for on-line measurement of APD parameters, 
TCP-IP/GPIB and TCP-IP/RS-232 converters for the data transfer and  experiment operation   from the control room.
PIN diode BPW34 used for neutron flux measurement and APD sample is placed at the distance of $\sim$3m 
from the neutron source to achieve the minimal possible intensity of neutron beam for the irradiation.
Other equipment is placed behind the concrete wall and connected  by Ethernet  to the computer in the
control room.

Three types of APD produced by Zecotek \cite{zecotek}, Ketek \cite{ketek}
and Hamamatsu \cite{hamamatsu} were investigated to understand 
dependences of APDs radiation hardness on the manufacturing technology. These APDs
were chosen as they are widely applied in nuclear and particle physics. The 
operational voltage and fluence equivalent to 1~MeV neutrons for  for these types of APDs are given
in table \ref{table:apd-cond}.
 
\begin{table}[hbt!]
\label{table:apd-cond}
\caption{The operational voltage and 1~MeV neutron fluence for different
irradiated APDs.} 
\vspace{5mm}
\centering
\begin{tabular}{|c|c|c|c|}
\hline
APD type & ref.&  $V_{bias}$,    &  1~MeV neutron   \\
         &    &[V]            &  fluence,  [n/cm$^2$] \\
\hline
Zecotek MAPD-3N       & \cite{zecotek}& 88.5       & 3.4$\pm$0.2$\cdot$10$^{12}$ \\
Ketek PM3350          & \cite{ketek}      & 23.5       & 2.5$\pm$0.2$\cdot$10$^{12}$ \\
Hamamatsu S12572-010P &  \cite{hamamatsu} & 69.2       & 6.5$\pm$0.6$\cdot$10$^{10}$ \\
\hline
\end{tabular}
\end{table}

Zecotek  MAPD-3N,  Ketek PM3350  and Hamamatsu S12572-010P were irradiated with the 1~MeV neutron doses of   3.4$\pm$0.2$\cdot$10$^{12}$ n/cm$^2$, 2.5$\pm$0.2$\cdot$10$^{12}$ n/cm$^2$ and  6.5$\pm$0.6$\cdot$10$^{10}$ n/cm$^2$, respectively. Doses were measured by the special PIN diode calibrated for a 1~MeV neutrons equivalent dose; the temperature during the irradiation and measurements was 
22$\pm$0.5$^\circ$C \cite{kerma}.

\section{\label{sec:measur} Results for irradiated APDs}

The tests of APDs before and after irradiation were performed using the photons from 
Light Emitted Diode (LED) and cosmic muons.  LED allows to investigate APD properties after 
irradiation  in single-photon mode of operation when signal to noise ratio is very low,  
in particular, the threshold variation of APD photon detection. The cosmic rays provide the 
possibility to study APDs with minimal ionizing particles (MIPs).   

\begin{figure}[hbtp]
 \centering
  \resizebox{7cm}{!}{\includegraphics{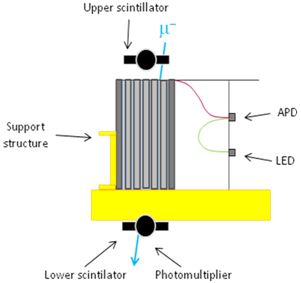}}
\caption{Setup for the APDs tests with cosmic muons and LED.}
\label{fig:fig2}
\end{figure}
 
For these purposes  the experimental setup was arranged as shown in Fig.\ref{fig:fig2}
The investigated APD was connected to the scintillators of one PSD module section \cite{psd} 
or to LED via an optical fibers. The cosmic muons penetrate the PSD scintillators with path length in range 16-200 mm depending on their declination angle.  
The coincidence of the signals from two scintillation counters placed upper and down 
the PSD module section provided a trigger for a DAQ system with frequency of about 10 counts per min. 
The Voltcraft PPS-12008 power supply was used as a HV supply for the MAPD optical sensor. The signal from the MAPD was processed by a fast amplifier and the resulting pulse-height distribution was collected by the Rohde \& Schwarz RTO1024 oscilloscope with 2GHz bandwidth.  

\begin{figure}[hbtp]
 \centering
  \resizebox{11cm}{!}{\includegraphics{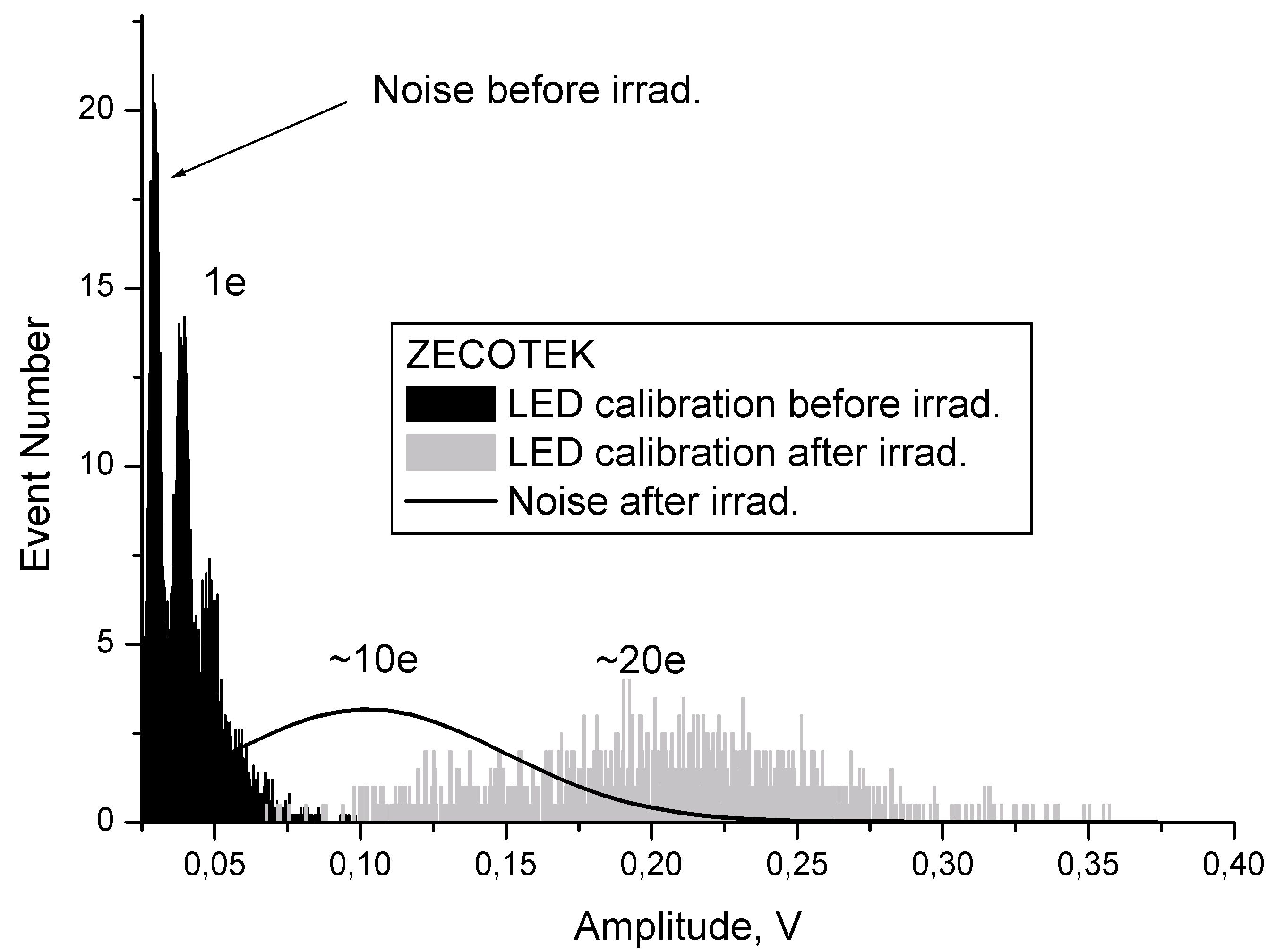}}
\caption{Test results of Zecotek
MAPD-3N with LED.}
\label{fig:fig3}
\end{figure}

The APD characteristics were measured before and after
the irradiation. The Capacitance-Voltage (C-V), Current-
Voltage (I-V), and Capacitance-Frequency (C-F) characteristics 
were studied using a dedicated testing setup at NPI  in 
\v Re\v z \cite{vasja_nim}.
After irradiation, the C-V technique showed significant decrease of 
hysteresis and fast but not complete self-annealing. 
The I-V curve revealed about 10$^3$ times increase of dark current after irradiation. 
The C-F study showed significant increase of short-living traps in Silicon.
The test results suggest an increase of internal APD noise,
especially of the high frequency, which depends on the
amount of short-living traps in the APD volume.

\begin{figure}[hbtp]
 \centering
  \resizebox{11cm}{!}{\includegraphics{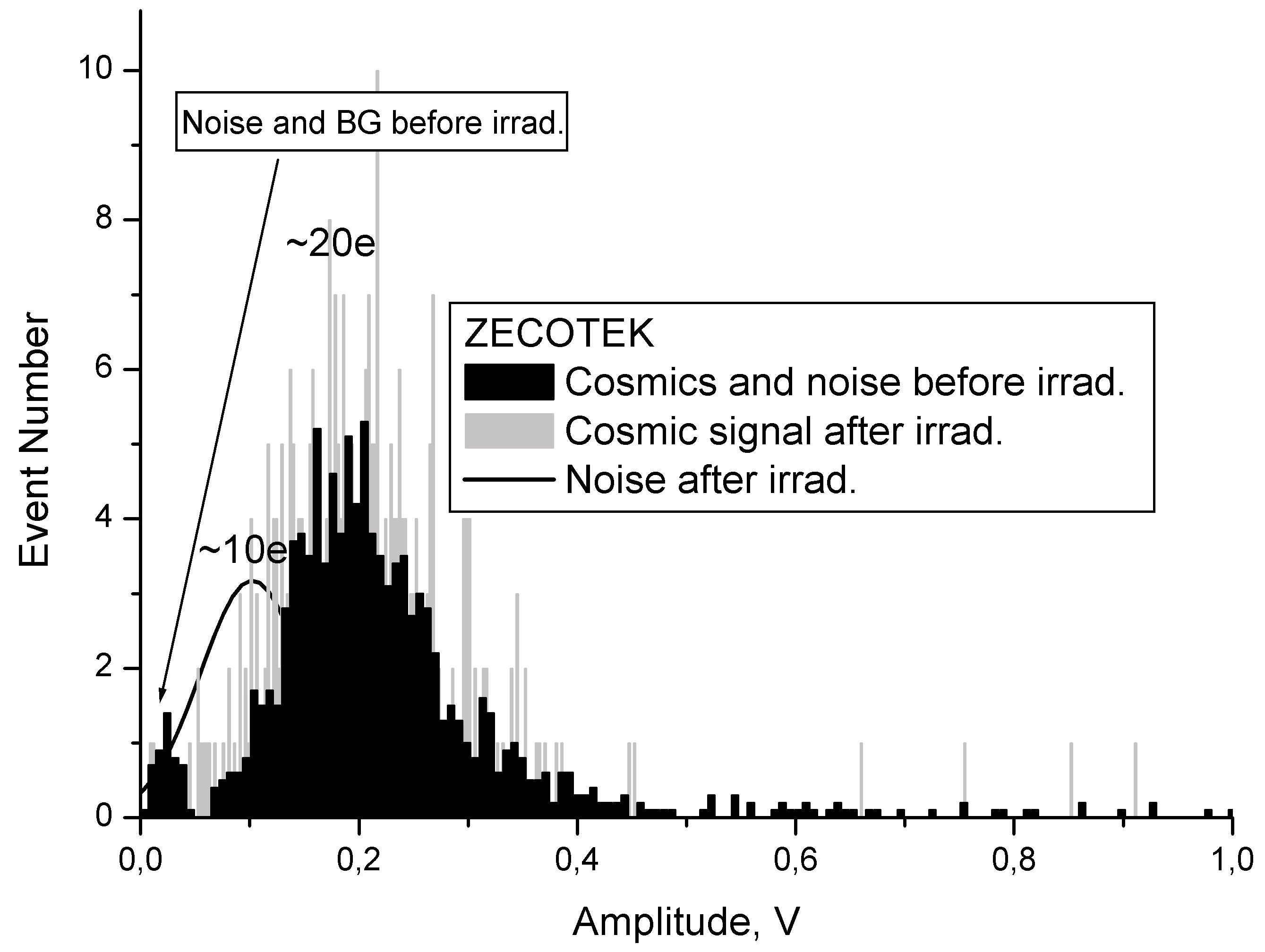}}
\caption{Test results of Zecotek
MAPD-3N with cosmic muons.}
\label{fig:fig4}
\end{figure}
 
The results of the Zecotek MAPD-3N studies with LED and with cosmic muons are presented
in Fig.\ref{fig:fig3} and Fig.\ref{fig:fig4}, respectively.  The dark and grey histograms
represent the APD amplitudes before and after irradiation, respectively.
The solid lines are the results of the noise signal shape approximation after irradiation.

Fig.\ref{fig:fig3} demonstrates  clear single and double photons peaks before irradiation.  
After irradiation   APD is unable to resolve single photons
due to high noise level ($\sim$10 p.e.). The amplitude from cosmic muons is defined by
the PSD prototype design and efficiency of the light collection.  
Fig.\ref{fig:fig4} shows that the averaged value of the signal amplitude from  Zecotek MAPD-3N
is $\sim$0.2~V corresponding to $\sim$20 p.e.  
The typical noise signal amplitude is $\sim$3 p.e. and $\sim$10 p.e. before and after irradiation,
respectively.  One can conclude that
the signal from APD does not change   drastically
and it is still well separated from the noise after irradiation.

\begin{figure}[hbtp]
 \centering
  \resizebox{11cm}{!}{\includegraphics{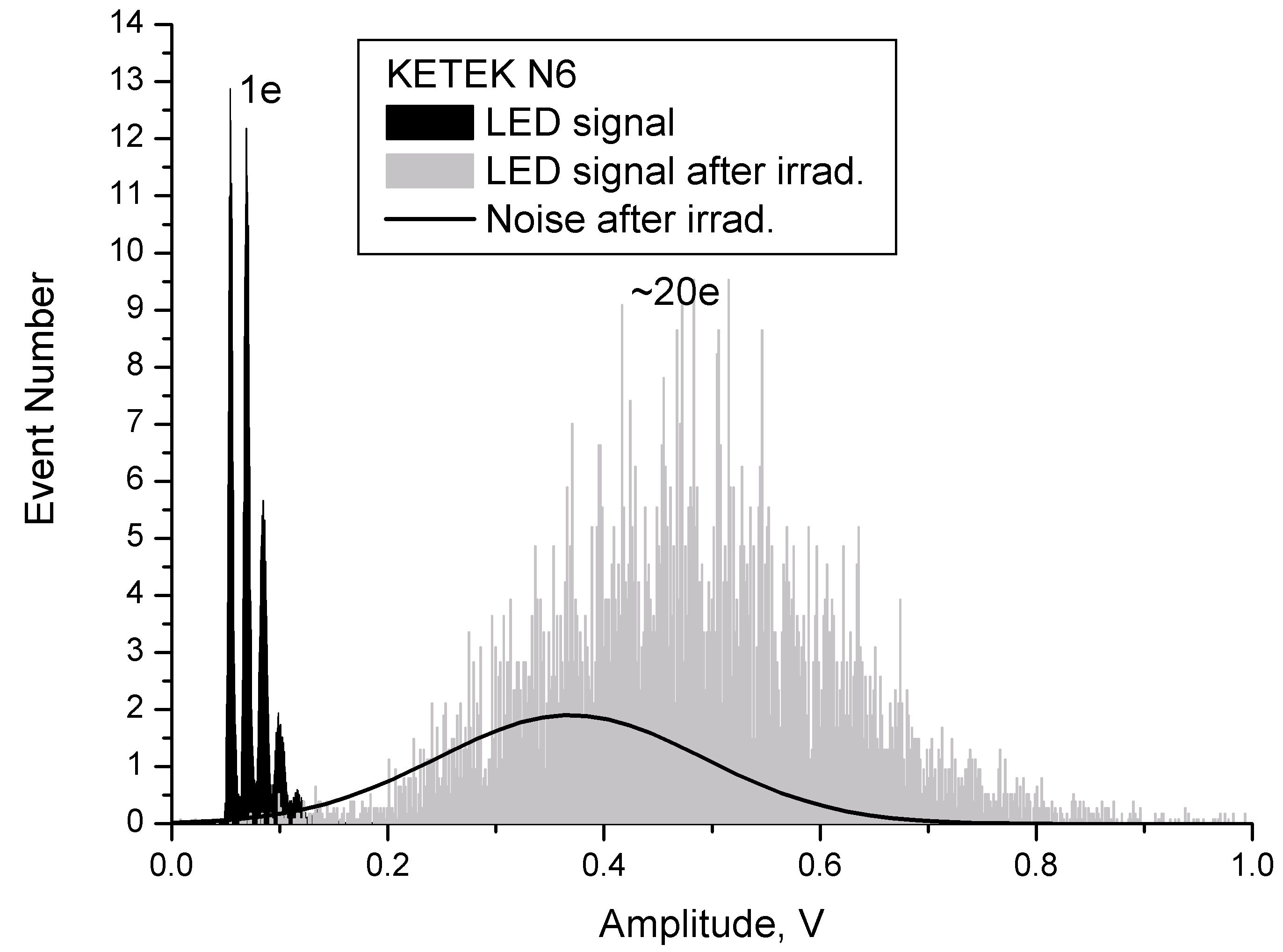}}
\caption{Test results of Ketek PM3350 with LED.}
\label{fig:fig5}
\end{figure}

\begin{figure}[hbtp]
 \centering
  \resizebox{11cm}{!}{\includegraphics{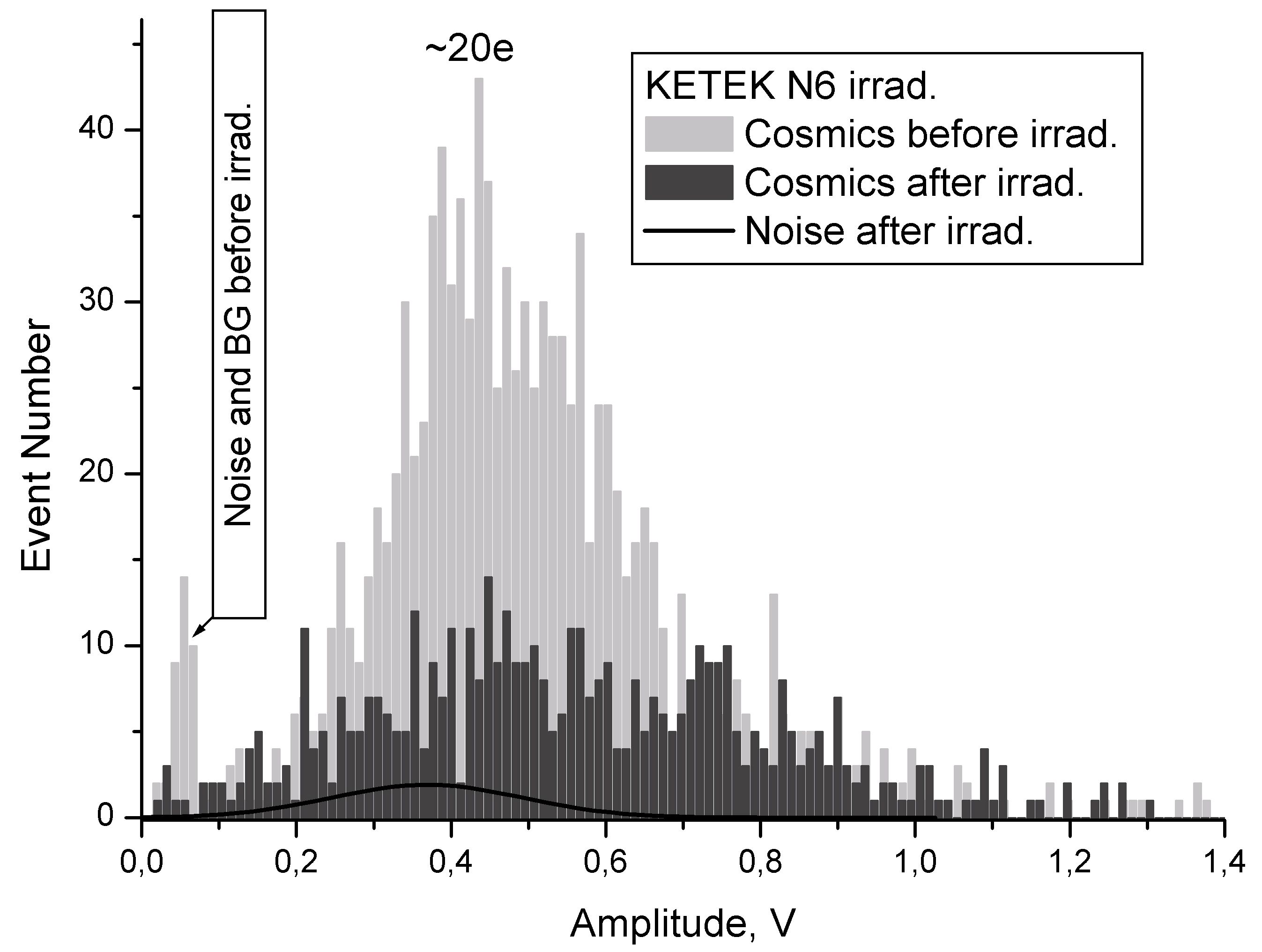}}
\caption{Test results of Ketek PM3350 with cosmic muons.}
\label{fig:fig6}
\end{figure}

The results of the Ketek PM3350 studies with LED and with cosmic muons are shown
in Fig.\ref{fig:fig5} and Fig.\ref{fig:fig6}, respectively.  The dark and grey histograms
represent the PM3350 amplitudes before and after irradiation, respectively.
The solid lines are the results of the noise signal shape approximation after irradiation.

The dark histogram shown in Fig.\ref{fig:fig5} demonstrates clear single and double photon peaks before irradiation. PM3350 after irradiation is unable to resolve single photons
due to high noise level which is $\sim$15 p.e. Fig.\ref{fig:fig6} shows that the averaged value of the signal amplitude from  Ketek PM3350 
is $\sim$0.4~V corresponding to $\sim$20 p.e.  
The signal
and noise peaks for irradiated  Ketek PM3350 are very close
which makes signal from noise separation difficult.
 
\begin{figure}[hbtp]
 \centering
  \resizebox{11cm}{!}{\includegraphics{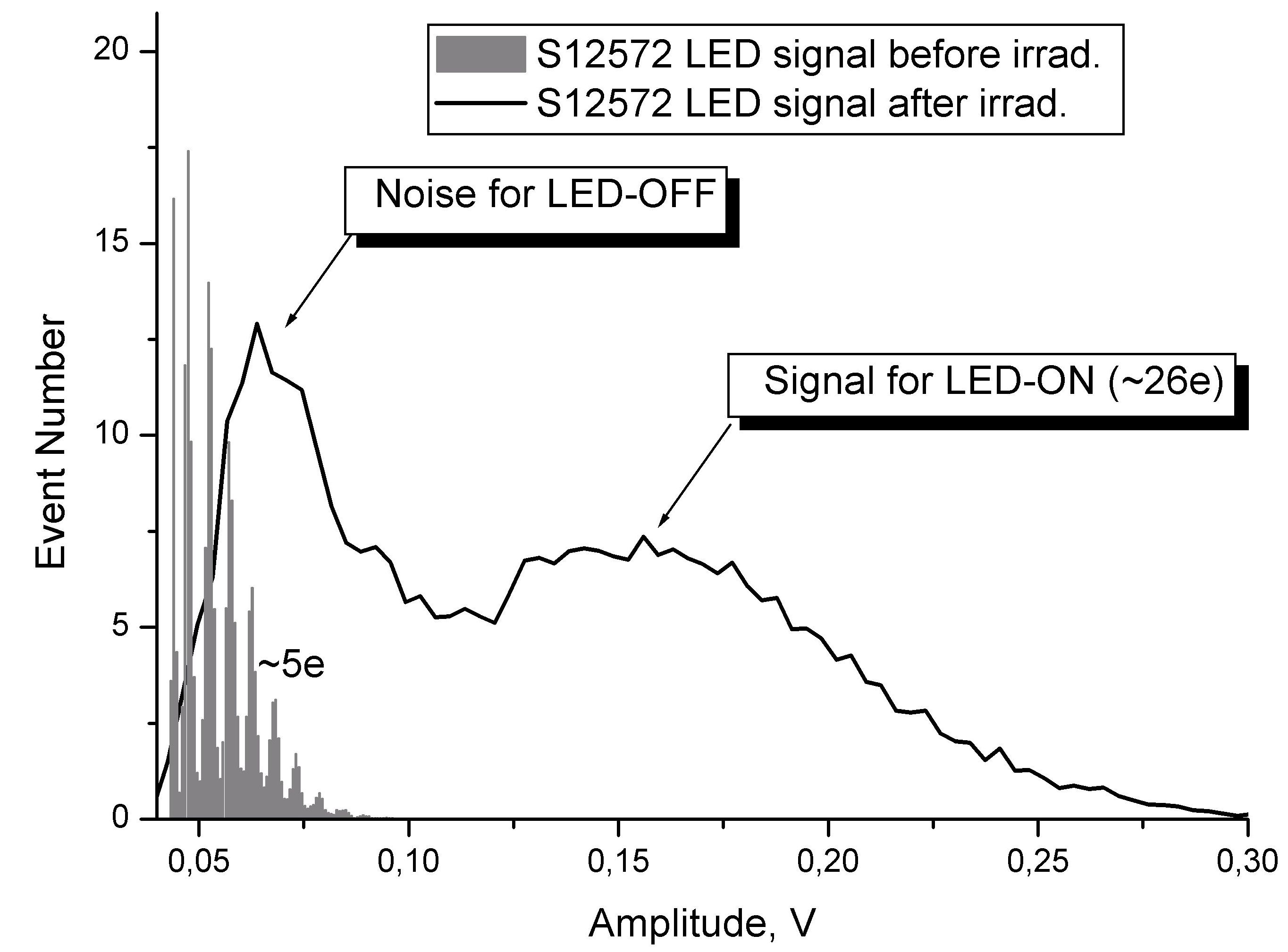}}
\caption{Test results of Hamamatsu  S12572-010P with LED.}
\label{fig:fig7}
\end{figure}

\begin{figure}[hbtp]
 \centering
  \resizebox{11cm}{!}{\includegraphics{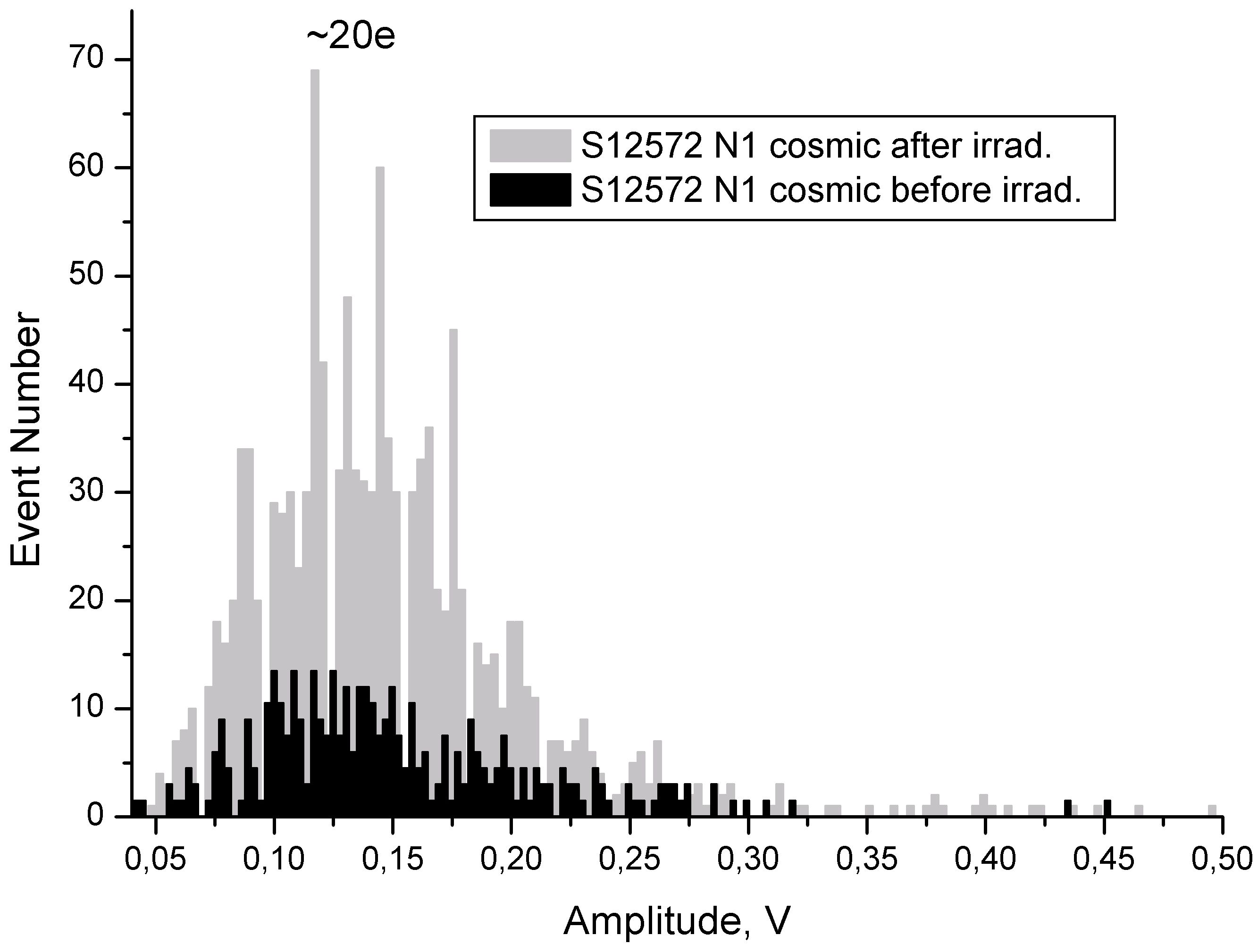}}
\caption{Test results of Hamamatsu  S12572-010P with cosmic muons.}
\label{fig:fig8}
\end{figure}

The results of the  Hamamatsu  S12572-010P studies with LED  
are demonstrated 
in Fig.\ref{fig:fig7}.  
The dark histogram and solid line 
represent the Hamamatsu S12572-010P amplitudes before and after irradiation, respectively.
One can see the good separation of the single photon peak before the irradiation. 
The averaged   value of the signal amplitude from  Hamamatsu S12572-010P 
after irradiation is $\sim$0.18 V corresponding to $\sim$26 p.e.  It is well separated from the noise peak 
($\sim$6 p.e.).

The Hamamatsu  S12572-010P  signal amplitudes  obtained with cosmic muons are
presented in Fig.\ref{fig:fig8}.  
The dark and grey histograms
represent the Hamamatsu  S12572-010P  amplitudes before and after irradiation, respectively.
The  averaged value of the signal amplitude from   Hamamatsu  S12572-010P 
is $\sim$0.12 V corresponding to $\sim$20 p.e.  
The signal from APD does not change drastically and it is still well separated from the noise after irradiation. However, one has to note that the neutron fluence for  Hamamatsu  S12572-010P 
was 30-50 times less than for Zecotek MAPD-3N and Ketek PM3350 (see table~1).
It will be necessary to perform the studies for the APDs as a function of the irradiation dose.

\section{\label{sec:conclusions} Conclusions}

\begin{itemize}
\item
The studies of the Zecotek MAPD-3N \cite{zecotek}, Ketek PM3350 \cite{ketek} and Hamamatsu  S12572-010P  \cite{hamamatsu} properties have been performed before and after irradition by neutrons  from cyclotron facility U120M at NPI of ASCR in 
\v Re\v z.  
\item 
It is demonstrated that the irradiation increase the APDs’ internal noise what leads to
inability to  detect single photons. 
\item
It is shown that the signal and noise peaks are well separated  for  Zecotek MAPD-3N and 
Hamamatsu  S12572-010P after irradiation. 
The Ketek PM3350 is unable to  separate the noise and  signal peaks for the
current version of the PSD module.   
\item
The obtained results are certainly important to design the detectors with APD light readout
for FAIR and NICA experiments. 
\item
The next steps will be a study of the radiation hardness of Ketek, Zecotek and Hamamatsu APDs with online dose monitoring, long time cosmic tests for all types of APDs and investigation of dependence of optimal avalanche amplification on absorbed radiation dose.
\end{itemize}
 \vspace{0.5cm}
The authors thank the NPI cyclotron and neutron generators staff for
excellent beam conditions and service, Z. Sadygov andcollaborants from Joint Institute for Nuclear Research (Dubna, Russia) for Zecotek APDs, and F. Guber and collaborants from Institute for Nuclear Research (Moscow, Russia) for help with the cosmics setup. This work was supported
by the European Community FP7 – Capacities, contract HadronPhysics3 no.283286, 
LG12007 of the Ministry of Education of the Czech
Republic and M100481202 of the Academy of Sciences of the Czech
Republic grants, by project MSMT LG14004 of Cooperation Program between JINR and NPI in 2013-2014 and RFBR 
under grant $N^o$13-02-00101a.

\begin{sloppypar}

\end{sloppypar}
\end{document}